\documentclass[a4paper,11pt]{article}
\usepackage{pos}

\newcommand{\LA}{\left\langle}
\newcommand{\RA}{\right\rangle}

\usepackage{lipsum}
\usepackage{titlesec}

\titlespacing\section{0pt}{10pt plus 0pt minus 0pt}{10pt plus 0pt minus 0pt}
\titlespacing\subsection{0pt}{8pt plus 0pt minus 0pt}{8pt plus 0pt minus 0pt}

\title{A new way to resum Lattice QCD equation of state at finite chemical potential}

\author*[a]{Sabarnya Mitra}
\author[a]{Prasad Hegde}
\author[b]{Christian Schmidt}

\affiliation[a]{Centre for High Energy Physics, Indian Institute of Science, 
   Bangalore - 560012, India}

\affiliation[b]{Fakult\"at f\"ur Physik, Universit\"at Bielefeld, Bielefeld D-33615, Germany}

\emailAdd{sabarnyam@iisc.ac.in}
\emailAdd{prasadhegde@iisc.ac.in}
\emailAdd{schmidt@physik.uni-bielefeld.de}

\abstract{
The Taylor expansion of thermodynamic observables at a finite baryon chemical potential $\mu_B$ is an oft-used method to circumvent the well-known sign problem of Lattice QCD. A reliable Taylor estimate demands sufficiently high-ordered calculations in chemical potential $\mu$ for a proper estimate of its radius of convergence. Owing to the associated difficulty and limitations of precision in calculating these high-order Taylor coefficients, it becomes essential to look for various alternative resummation schemes which can work around this computational hurdle. Recently, a way to resum exponentially, the contributions of the first $N$ baryon charge density correlation functions $D_1,\dots,D_N$ to the Taylor series to all orders in $\mu_B$ was proposed in Phys. Rev. Lett. 128, 2, 022001 (2022). Since the correlation functions $D_n$ are calculated stochastically using estimates from different random volume sources, the resummation formulation gets affected by biased estimates, which can become very drastic and can radically misdirect the calculations for large values of $N$, $\mu$ and also higher order $\mu$ derivatives of free energy.
In this work, we present a cumulant expansion procedure that allows to investigate and regulate these biased estimates at different orders in $\mu$. We find that the unbiased estimates in the cumulant expansion can truly capture the genuine higher-order stochastic fluctuations of the higher order correlation functions, which got suppressed by the exponential resummation formulation. Finally, we introduce an unbiased formalism of exponential resummation, which when expanded in a series, can exactly reproduce the Taylor series upto a desired order in $\mu$. This allows to regain the knowledge of reweighting factor and many other important properties of the partition function, which got entirely lost while implementing the cumulant expansion scheme.}

\FullConference{%
   The 39th International Symposium on Lattice Field Theory (Lattice 2022) \\
   8-13 August 2022 \\
   H\"{o}rsaalzentrum Poppelsdorf, Universit\"{a}t Bonn, Bonn, Germany 
}

\begin{document}
\maketitle

\section{Introduction}

The QCD Equation of State (EoS), illustrating the QCD Phase diagram is of significant importance in the parlance of QCD phase transitions and also in the study of heavy-ion collisions~\cite{Bernhard:2016tnd,Parotto:2018pwx,Monnai:2019hkn,Everett:2020xug}.
In principle, the entire phase diagram can be completely explained from a comprehensive study of the gauge theory of QCD. But, in reality, it is still a conjecture from a practical standpoint, with many salient and robust features remaining to be established. Hence, for a proper unfazed conclusion, an unambiguous thermodynamic approach is adopted, which revolves around the important calculations of the estimates of various thermodynamic observables. 

The system considered, resembles a grand canonical ensemble of quarks interacting via gluons, described by a grand canonical partition function $\mathcal{Z}$($\mu,V,T$), which, in principle, is given as a path integral over all the constituent particle (quark) and gauge field (gluon) configurations. Unfortunately, this path integral formulation yields an intractable, infinite-dimensional integral. Although lattice QCD averts this problem by rendering this integral to a finite-dimensional one, the complex integral measure at a finite $\mu$ inhibits the implementation of Monte-Carlo importance sampling (MCIS). By virtue of the reweighting procedure~\cite{Fodor:2002fdr,Fodor:2004jhe, Ejiri:2004prd, Saito:2014prd}, although the measure being weighted at zero $\mu$ becomes real, the complex measure problem assumes the form of the sign problem~\cite{Nagata:2021nag,Aarts:2014cpod,deForcrand:2009zkb}, which manifest in the observable part of the integral. On a positive note, reweighting enables the application of MCIS for calculating $\mathcal{Z}$ by making the integral measure semi-positive definite. 

The Taylor expansion of thermodynamic observables upto the first $N$ coefficients~\cite{Bazavov:2017dus,Bollweg:2021vqf} as a function of $\mu$ is one of the numerous methods~\cite{Aarts:2009yj,Cristoforetti:2012su,Aarts:2013uxa,Sexty:2013ica,Fukuma:2019uot,Borsanyi:2018grb,Ratti:2018ksb} adopted to evade the sign problem in Lattice QCD. The slow rate of convergence and non-monotonic behaviour of the Taylor series for a wide range of temperatures necessitate computations upto sufficiently high orders in $\mu$, invoking calculation of higher-order Taylor coefficients. This directs one towards resummation of Taylor series~\cite{Gavai:2008zr, Dimopoulos:2021vrk, Nicotra:2021pos, Giordano:2019slo,Allton:2005gk,Gavai:2003zr}, which allows to conduct an all-ordered calculation with the knowledge of a few Taylor coefficients. The exponential resummation~\cite{Mondal:2021jxk} is one such resummation method, instrumental in our work. 

In this work, we present the mathematical form of Taylor expansion and exponential resummation. We then comprehensively discuss about the emergence of biased estimates in exponential resummation, which has the potential to become highly problematic in the regime of large values and higher orders of $\mu$. We then come across the formulation of cumulant expansion~\cite{Mitra:2022prd,doi:10.1143/JPSJ.17.1100, Endres:2011jm,Ejiri:2008prd,Ejiri:2010prd}, which allows an order-by-order analysis of biased estimates, but unfortunately at the cost of the reweighting factor and hence, the invaluable partition function itself. Finally, we present an unbiased formulation of exponential resummation, which reproduces the Taylor (QNS) expansion upto a given order of $\mu$ apart from a newly defined reweighting factor and partition function altogether. 

\section{Setup of the simulation}
\label{sec:setup}

 In our work, we have used Highly Improved Staggered Quark (HISQ) action~\cite{Follana:2006rc,Bazavov:2011nk,Bazavov:2014pvz} for the fermions and tree-level improved Symanzik gauge action~\cite{Symanzik:1983nucl,Karsch:2000nucl} for the gauge fields. The work has been done on a $32^3 \times 8$ lattice, using 2+1 flavor QCD with the quark masses chosen to satisfy $m_u = m_d = m_s/27$. With a fixed lattice spacing and coupling parameter $\beta$, these masses are tuned appropriately to their physical values, so that they produce physical pion and kaon masses, as directed by chiral perturbation theory. This therefore fixes the line of constant physics for our work~\cite{Bazavov:2017dus, Bazavov:2018mes, Bollweg:2022rps}. 
 We have collected gauge configurations for two temperatures at T $= 135$ and $157$ MeV, which in $\beta$ scale, corresponds to $\beta = 6.245$ and $6.390$ respectively. We have worked with 20K configurations for both baryon ($\mu_B$) and isospin ($\mu_I$) chemical potentials. Recent work for $176$ MeV is in progress and also the number of gauge field configurations is increased for $\mu_B$ for more statistics. Although we worked mostly upto $D_4$, all the eight derivatives till $D_8$ are calculated stochastically using $\mathcal{O}(500)$ random volume sources (RVS) per configuration. All these correlation functions $D_n$ for $n \leq N$ can be expressed as different linear combinations of traces~\cite{Allton:2005gk,Gavai:2004sd}, involving products of fermion propagator $\mathcal{M}^{-1}$ and different ordered $\mu$ derivatives of fermion matrix $\mathcal{M}$. 
The stochastic calculation of these traces arises due to the inexact computation of $\mathcal{M}^{-1}$.
 A detailed description of the gauge ensembles and scale setting 
can be found in Ref.~\cite{Bollweg:2021vqf}.

 \section{Taylor series and Exponential Resummation}

   \hspace{4.5mm} The Taylor expansion of excess pressure $\Delta P/T^4 = P(\mu,T)/T^4 - P(0,T)/T^4$ and number density $\mathcal{N}/T^3$, in terms of $\mu_B$ upto the first $N$ derivatives~\cite{Bazavov:2017dus,Bollweg:2021vqf} are given by
   
        \begin{equation} 
            \frac{\Delta P_N^Q(T,\mu_B)}{T^4} = \frac{1}{VT^3} \ln  \bigg[\frac{\mathcal{Z(\mu)}}{\mathcal{Z}(0)}\bigg] = \sum_{n=1}^N \frac{X_{2n}}{(2n)!} \hspace{1mm}\hat{\mu}_B^{2n} 
          \label{QNS}
        \end{equation}

where $X_{n}$ are the $n$th order quark number susceptibilities (QNS) and $\hat{\mu}_B = \mu_B/T$. The CP symmetry of QCD~\cite{Allton:2005gk} ensures that pressure and number density constitutes even and odd series in $\mu_B$ respectively.  The exponentially resummed estimate of excess pressure is given by

\begin{equation}
\begin{split}
\frac{\Delta P^R_N(T,\mu_B)}{T^4}
&= \frac{1}{VT^3}\ln \Bigg \langle \text{Re} \Bigg[ \exp \left(\frac{1}{4}\hspace{.5mm}\sum_{n=1}^N \overline{D}_n(T) \hspace{1mm}\hat{\mu}_B^n \right) \Bigg]\Bigg \rangle, \hspace{2mm}
  \overline{D}_n(T) = \frac{1}{N_{R}} \sum_{r=1}^{N_{R}} \Tilde{D}_n^{(r)}(T)\\
\end{split}
\label{eq:Resum pressure}
\end{equation}

\begin{equation*}
 \begin{aligned}
    \Tilde{D}_n^{(r)}(T) = \frac{D_n^{(r)}(T)}{n!}  & & \textnormal{where}\ \hspace{1.5mm}D_n^{(r)}(T)
= \frac{\partial^n}{\partial \hat{\mu}_B^n} \ln\big[\det M^{(r)}\left(T,\mu_B\right)\big]\Big|_{\mu_B=0}
\end{aligned}
\end{equation*}

The corresponding resummed and Taylor estimate of number density is given by 

\begin{equation}
    \frac{\mathcal{N}_N^{R,Q}}{T^3} = \frac{\partial}{\partial \hat{\mu}_B}\bigg[\frac{\Delta P_N^{R,Q}(T,\mu_B)}{T^4}\bigg]
\end{equation}

\vspace{3mm}

The factor of \hspace{.4mm}$1/4$, as mentioned in Eqn.~\eqref{eq:Resum pressure}, represents staggered signature of fermion action. Also in this equation, the angular brackets $\langle . \rangle$ represent the gauge ensemble average and $D_n^{(r)}$ is the $r$th estimate of the derivative $D_n$.  In continuum limit, these derivatives are the integrated $n$-point correlation functions of the product of the zeroth component of the four baryon current density $J_{\alpha}$ = $(J_0,\Vec{\mathbf{J}}\hspace{.3mm})$ at different spacetime points $x$, which are given by $D_n = \int d^4x_1 \hspace{.8mm}d^4x_2\hspace{.8mm} ...\hspace{.8mm} d^4x_n \hspace{.8mm}J_0(x_1)\hspace{.8mm}J_0(x_2)\hspace{.8mm}...\hspace{.8mm}J_0(x_n)$~\cite{Mondal:2021jxk}. The \hspace{.5mm}CP\hspace{.5mm} symmetry ensures real-valued $\mathcal{Z}$, thereby dictating that all $D_n$ are real for even $n$ and imaginary for odd $n$. causing only the real part of the exponential to be considered as shown in eqn.~\eqref{eq:Resum pressure}. It is therefore evident that
\begin{equation}
\Delta P_N^R/T^4 = \Delta P_N^Q/T^4 + 
\sum_{n>N}^\infty 
\Big \langle (\overline{D}_1)^{A_1} (\overline{D}_2)^{A_2} \dots (\overline{D}_N)^{A_N} \Big \rangle \; \hat{\mu}_B^n
\label{eq:RQ}
\end{equation} 
where every $k$-point correlation function $D_k$ satisfies $\sum_{k=1}^N k \cdot A_k = n$. The number density $\mathcal{N}_N/T^3$ also exhibits similar comparative behaviour resembling eqn.~\eqref{eq:RQ}. 


\section{Problem of Biased Estimates and Cumulant Expansion} 

\subsection{Biased estimates}

 \begin{figure}[h]
 \centering
\includegraphics[width=0.47\textwidth]{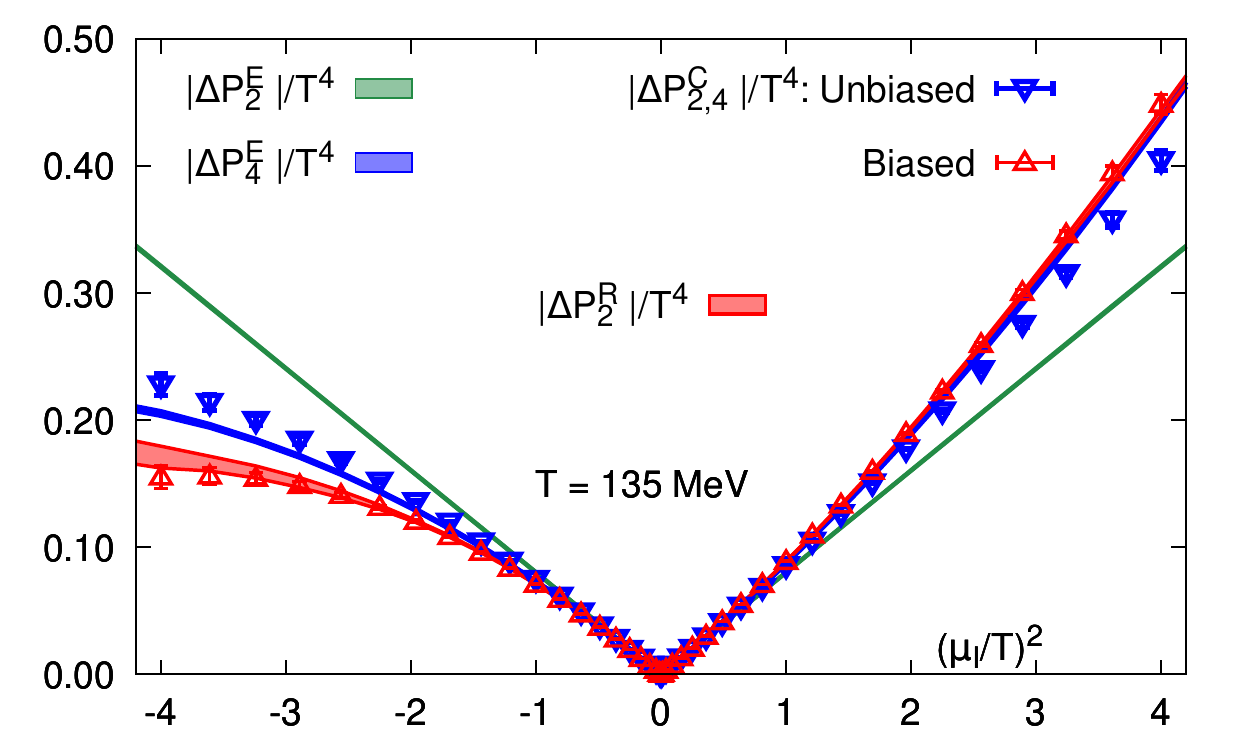} \hspace{0.01\textwidth}%
\includegraphics[width=0.47\textwidth]{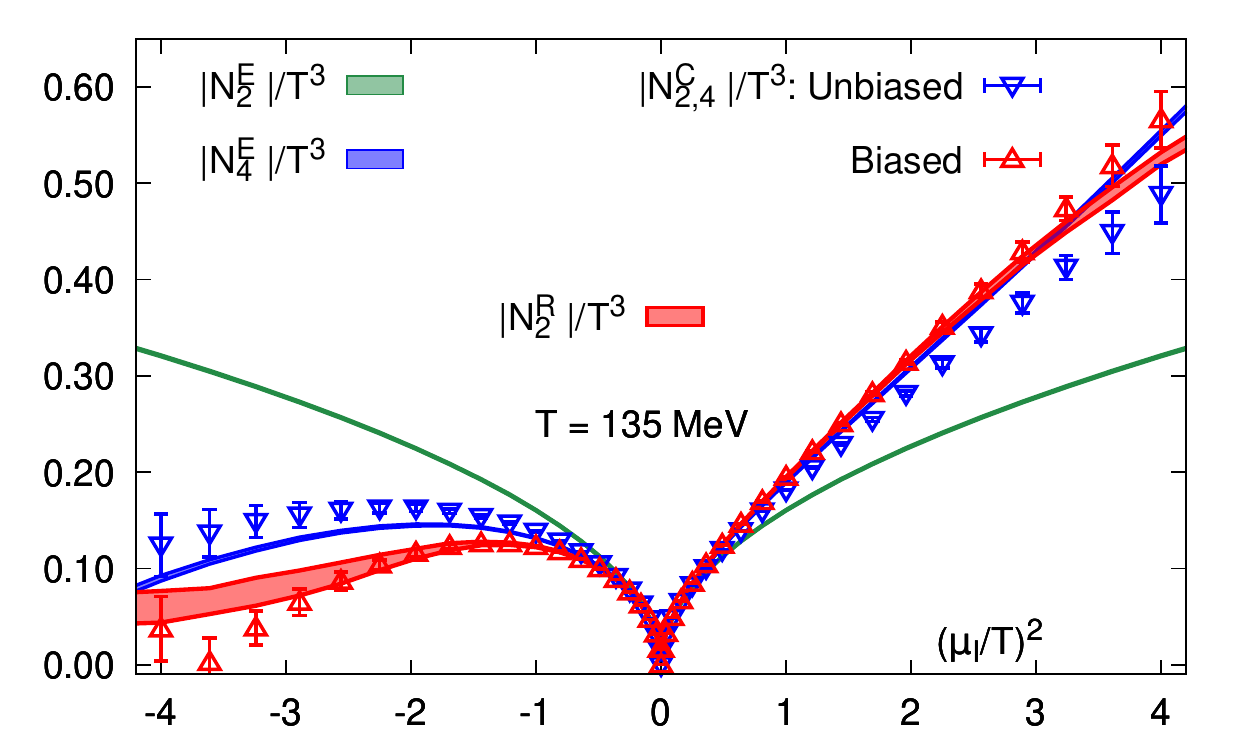}\\

\includegraphics[width=0.47\textwidth]{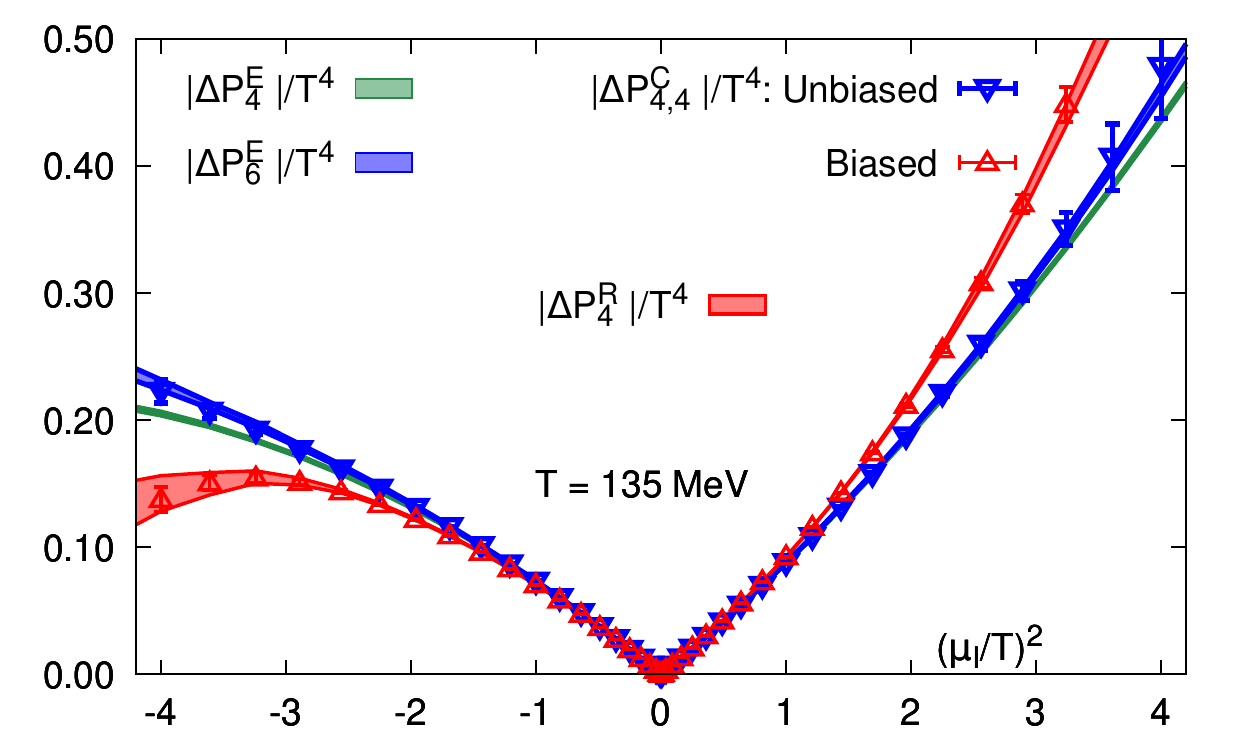} \hspace{0.01\textwidth}%
\includegraphics[width=0.47\textwidth]{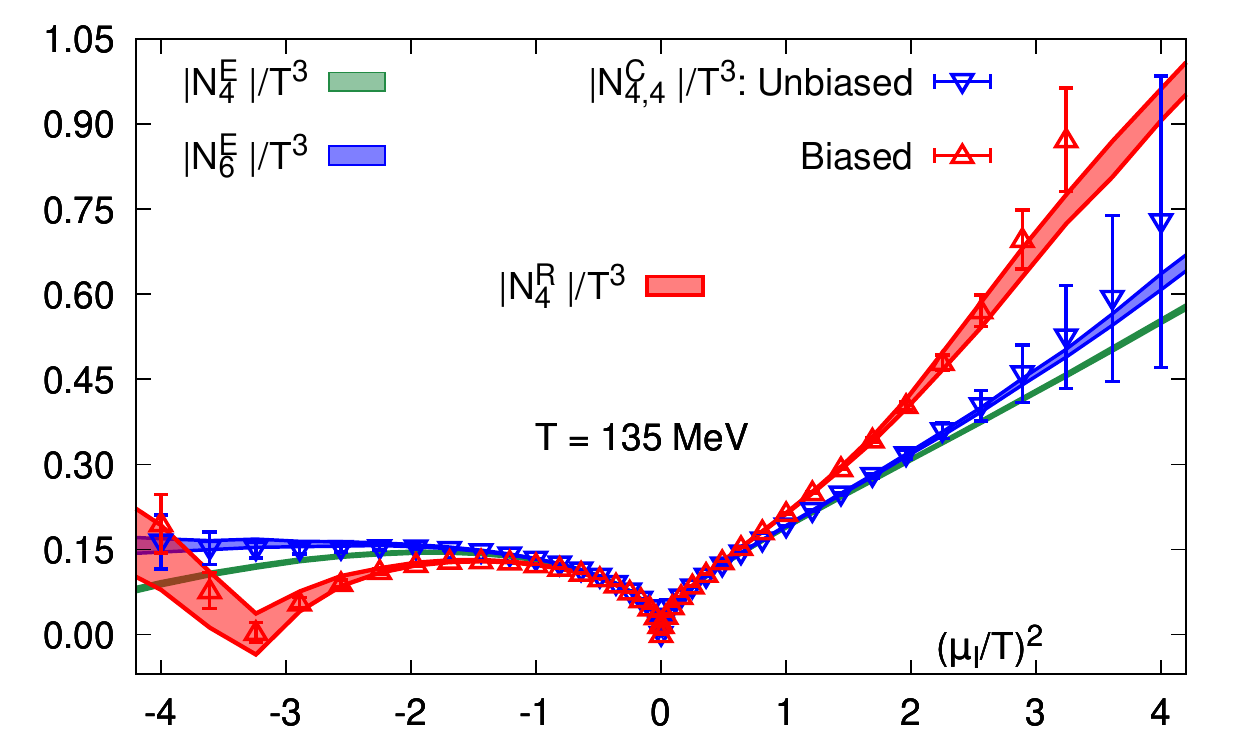}%
\caption{
     Isospin pressure (left) and number density (right) plots for T = 135 MeV. 
     }
\label{Fig: Isospin pressure}
\end{figure}

 As shown in Fig.~\ref{Fig: Isospin pressure} and also in~\cite{Mondal:2021jxk}, the resummed results differ appreciably from the QNS counterparts in the regime of higher values of $\mu_I$. This stark difference arises from the higher order contribution terms, as indicated in eqn.~\eqref{eq:RQ}. More significantly, as given in eqn.~\eqref{eq:Resum pressure}, the different powers of these different derivative estimates, from quadratic power onwards give rise to biased estimates. This is because, some given random vector estimates are raised to higher powers than the others, thereby treating different estimates on different footing in the sample of estimates.

 \begin{equation}  
    \begin{split}
        \big(\overline{D}_n\big)^m = \Bigg[\frac{1}{N_R} \sum_{r=1}^{N_R} D_n^{(r)}\Bigg]^m = 
    \Bigg[\bigg(\frac{1}{N_R}\bigg)^m \sum_{r_1=1}^{N_R}...\sum_{r_m=1}^{N_R} D_n^{(r_1)}...D_n^{(r_m)}\Bigg] \\
    \approx \text{Biased estimate} +
     \bigg(\frac{1}{N_R}\bigg)^m\mathop{\sum^{N_R}...\sum^{N_R}}_{r_1\neq ...\neq r_m} D_n^{(r_1)}...D_n^{(r_m)}
    \end{split}
    \end{equation}   

 The effects of these problematic biased estimates can become very pronounced and drastic, specially in the regime of large values and higher orders of $\mu_I$ and estimating observables which are higher order $\mu_I$ derivatives of free energy. This therefore motivates one to truncate the resummed series in terms of different powers of $\mu_I$ and analyse the biased estimates for different orders of $\mu_I$.
 
 \subsection{Cumulant Expansion}

 The cumulant expansion of eqn.~\eqref{eq:Resum pressure} upto $M$ cumulants in $\mu_I$ yield (barring the $1/VT^3$ factor)
   
  \begin{equation}  
            \ln \LA e^{X_N} \RA = \sum_{n=1}^M \frac{\kappa_n^N}{n!} + \mathcal{O}(\kappa_{M+1}^N)
            , \hspace{2mm} X_N = \sum_{n=1}^N \overline{D}_{2n}(T) \hspace{1mm}\hat{\mu}_I^{2n} 
        \label{Cumulant expansion}
        \end{equation}

    We exploited the efficacy of cumulant expansion for $\mu_I$, where there is no sign problem, because of the vanishing odd-ordered derivatives and also because of which, $X_N$ in eqn.~\eqref{Cumulant expansion} is manifestly real, ensuring $\mathcal{O}(10K)$ gauge configurations is good enough for an appreciable signal. We worked with only the first $M = 4$ cumulants and computed biased and unbiased cumulants, where the biased cumulants $\kappa_{b,n}^N$ are given by

    \begin{equation}   
        \begin{aligned}
    \kappa_{b,1}^N &= \LA X_N \RA \\
    \kappa_{b,2}^N &= \LA X_N^2 \RA - \LA X_N \RA^2 \\
    \kappa_{b,3}^N &= \LA X_N^3 \RA - 3\LA X_N^2 \RA \LA X_N \RA + 2\LA X_N\RA^3 \\
    \kappa_{b,4}^N &= \LA X_N^4 \RA - 4\LA X_N^3 \RA \LA X_N \RA + 12\LA X_N^2\RA \LA X_N\RA^2 - 6\LA X_N\RA^4 - 3\LA X_N^2\RA^2
    \label{eq:biased cumu}
        \end{aligned}
    \end{equation}
   
     For unbiased cumulants $\kappa_{u,n}^N$, we replace $X_N^n$ with $U_n[X_N]$ for each $n$, in the cumulants of eqn.~\eqref{eq:biased cumu}. Here $U_n[X_N]$ is the unbiased $n$th power of $X_N$, where $X_N = \sum_{n=1}^N \overline{D}_n \hspace{1mm}\hat{\mu}_B^n$ and unbiased $n$th power of $D_m$ is given by

     \begin{equation}
  U_n[D_m] = \frac{n!}{\prod_{k=0}^{n-1} (N_{R}-k)!}
  \mathop{\sum^{N_{R}}...\sum^{N_{R}}}_{r_1\neq ...\neq r_n} D_m^{(r_1)}...D_m^{(r_n)}
  \label{Unbiased formula of D}
  \end{equation}

 
    


As shown in the plots in Fig.~\ref{Fig: Isospin pressure}, the biased and unbiased results of pressure and number density are in good agreement with the resummed and QNS results of similar orders respectively. The unbiased cumulants managed to capture more higher-order fluctuations, which got suppressed by the exponential behaviour of the resummed series~\cite{Mitra:2022prd}. The unbiased cumulant expansion results hence, demonstrated that the difference between the resummed and QNS results is attributable to the difference between biased and unbiased estimates. 

But, while incorporating unbiasedness at different orders, the truncation of the resummed series led to the loss of the reweighting factor and partition function altogether. This inspired the idea of a newly defined exponential resummation scheme which would, in principle reproduce QNS upto the desired order in $\mu$. In addition, a numerically different partition function with an associated new reweighting factor is obtained, thereby re-enabling the essential calculations of phasefactor and roots of partition function.

\section{Unbiased Exponential Resummation}

  Motivated by the isospin results, we have implemented this new formalism of an unbiased exponential resummation using $\mu_B$. Unlike $\mu_I$, the odd-ordered derivatives are non-vanishing and imaginary for $\mu_B$ and hence, it is necessary to extract the real part following eqn.~\eqref{eq:Resum pressure} to obtain the expression of the partition function $\mathcal{Z}$. In this formalism, all mathematical manipulations are done at the level of individual RVS present within every gauge configuration constituting the gauge ensemble. We have worked in two bases, which are stated as follows:

\subsection{Chemical potential basis}

 In $\mu$ basis, with this new formalism, we define the unbiased pressure from a newly defined partition function following the usual prescription of the exponential resummation as follows:
        \begin{align}   
     \Delta P_{ub}^{N}(\mu) = \frac{1}{VT^3} \hspace{1mm}\ln \hspace{1mm} \mathcal{Z}_{ub}^{N}(\mu) , \hspace{4mm}\mathcal{Z}_{ub}^{N}(\mu) = \left \langle \text{Re}\bigg[\exp \Big(A_N(\mu)\Big)\bigg] \right \rangle , \hspace{4mm} 
    A_N(\mu) &= \sum_{n=1}^{N} \mu^n\hspace{.5mm}\frac{\mathcal{C}_{n}}{n!}
     \label{mu basis}
      \end{align}  
     
      where the $\mathcal{C}_n$ for $1 \leq n \leq 4$ are given as follows:

\begin{align}
    \mathcal{C}_1 &= \overline{D_1}, \notag \\
    \mathcal{C}_2 &= \overline{D_2} + \left(\overline{D_1^2} - \overline{D_1}^2\right), \notag \\
    \mathcal{C}_3 &= \overline{D_3} + 3\left(\overline{D_2D_1} - \overline{D_2}\;\overline{D_1}\right) + 
           \left(\overline{D_1^3} - 3\,\overline{D_1^2}\;\overline{D_1} + 2\,\overline{D_1}^3\right), \notag \\
    \mathcal{C}_4 &= \overline{D_4} + 3\left(\overline{D_2^2} - \overline{D_2}^2\right)+ 4\left(\overline{D_3D_1} - \overline{D_3}\;\overline{D_1}\right) +
      6\left( \overline{D_2D_1^2} - \overline{D_2}\;\overline{D_1^2}\right) \notag \\
   &- 12 \left(\overline{D_2D_1}\;\overline{D_1} - \overline{D_2}\;\overline{D_1}^2\right) + 
     \left(\overline{D_1^4} - 4\,\overline{D_1^3}\;\overline{D_1} + 12\,\overline{D_1^2}\;\overline{D_1}^2 - 6\,\overline{D_1}^4 - 3\,(\overline{D_1^2})^2\right) 
\label{mu coefficients}
\end{align}


 Here, the powers of different $D_n$ are the unbiased powers of the respective different ordered derivatives, calculated as per eqn.~\eqref{Unbiased formula of D}. The analysis from this basis is important in the sense, that the degree of the unbiasedness in $\mu$ is exactly identical with the degree of the polynomial $A(\mu)$ as given in eqn.~\eqref{mu basis}. This therefore ascertains the exact order of Taylor or QNS expansion in $\mu$, it will achieve, apart from the prescence of important beyond the QNS-order contributions, still comprising biased estimates.

 \subsection{Cumulant basis}

 In cumulant basis, a new variable $W$ is defined, where $W_N =\sum_{n=1}^N \frac{\mu^n}{n!}D_n$, we have
   
      \begin{align}   
     \Delta P_{ub}^{M}(W_N) = \frac{1}{VT^3} \hspace{1mm}\ln \hspace{1mm} \mathcal{Z}_{ub}^{M}(W_N) , \hspace{4mm}\mathcal{Z}_{ub}^{M}(W_N) = \left \langle \text{Re} \bigg[\exp \Big(Y_M(W_N)\Big)\bigg] \right \rangle , \hspace{4mm} 
    Y_M(W_N) = \sum_{n=1}^{M} \frac{\mathcal{L}_{n}(W_N)}{n!}
      \label{eq:cumulant basis}
      \end{align}

    which would reproduce exactly the first M cumulants in unbiased cumulant expansion of excess pressure. The different unbiased powers of derivatives are calculated as before, as given in eqn.~\eqref{Unbiased formula of D}. The $\mathcal{L}_n (W)$ of eqn.~\eqref{eq:cumulant basis} upto $M = 4$ for $1 \leq n \leq 4$ are as follows:

       \begin{equation}    
 \begin{aligned}
    \mathcal{L}_1(W) &= (\overline{W}) \\
    \mathcal{L}_2(W) &= \bigg[\Big(\overline{W^2}\Big) - \Big(\overline{W}\Big)^2\bigg] \\
    \mathcal{L}_3(W) &= \bigg[\Big(\overline{W^3}\Big) - 3 \hspace{1mm}\Big(\overline{W^2}\Big) \hspace{1mm}\Big(\overline{W}\Big) + 2 \hspace{1mm}\Big(\overline{W}\Big)^3\bigg] \\
    \mathcal{L}_4(W) &= \bigg[\Big(\overline{W^4}\Big) - 4 \hspace{1mm}\Big(\overline{W^3}\Big) \hspace{1mm}\Big(\overline{W}\Big) + 12 \hspace{1mm}\Big(\overline{W^2}\Big) \hspace{1mm}\Big(\overline{W}\Big)^2 - 6 \hspace{1mm}\Big(\overline{W}\Big)^4 - 3\hspace{1mm} \Big(\overline{W^2}\Big)^2\bigg]
 \end{aligned}
 \end{equation}

 
 


 
 
 \section{Results: Comparison between Biased and Unbiased formalism}

  The cumulant basis provides much more number of terms in addition to those of $\mu$ basis. Although $\mu$ basis is important for simplicity and first-hand understanding, the cumulant basis ensures a faster rate of convergence and agrees well with the $\mu$ basis results, as the additional terms get almost cancelled out among themselves. This also vindicates a genuine series expansion.
  The results in unbiased resummation are carried out therefore primarily, in cumulant basis.

 \begin{figure}[h]
 \centering
    \includegraphics[width=0.47\textwidth]{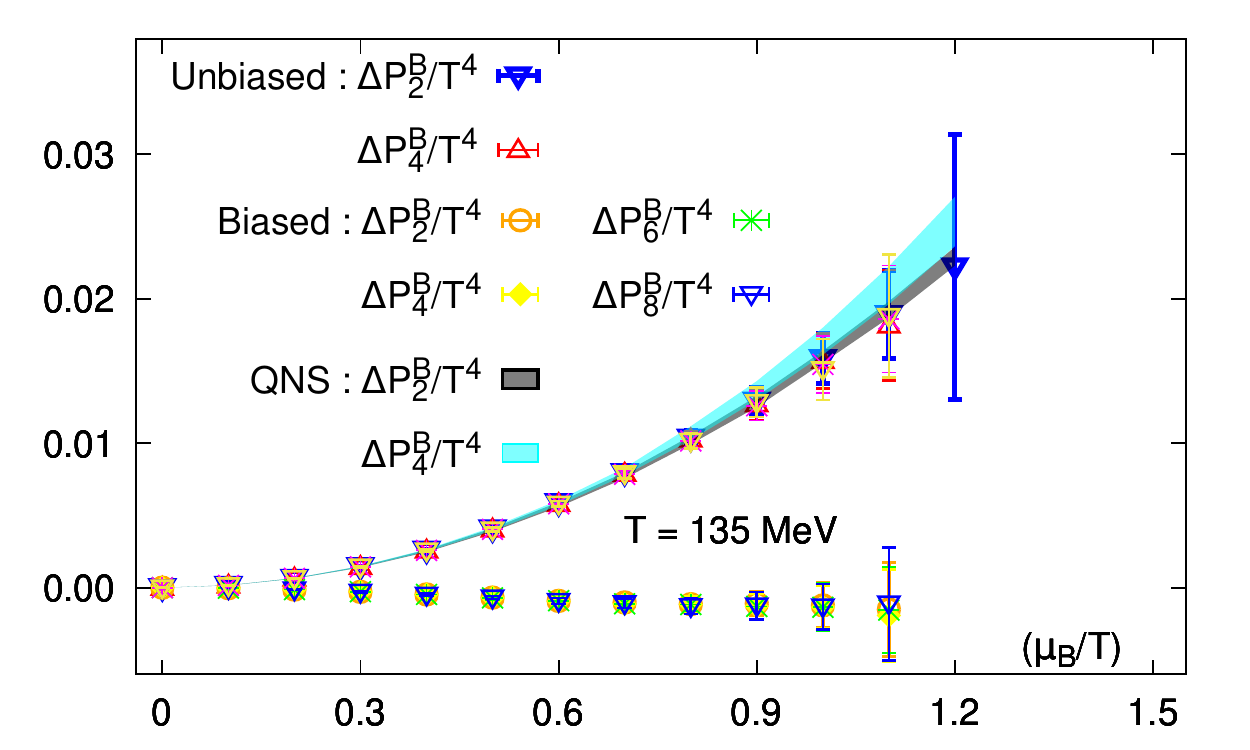} 
    \includegraphics[width=0.47\textwidth]{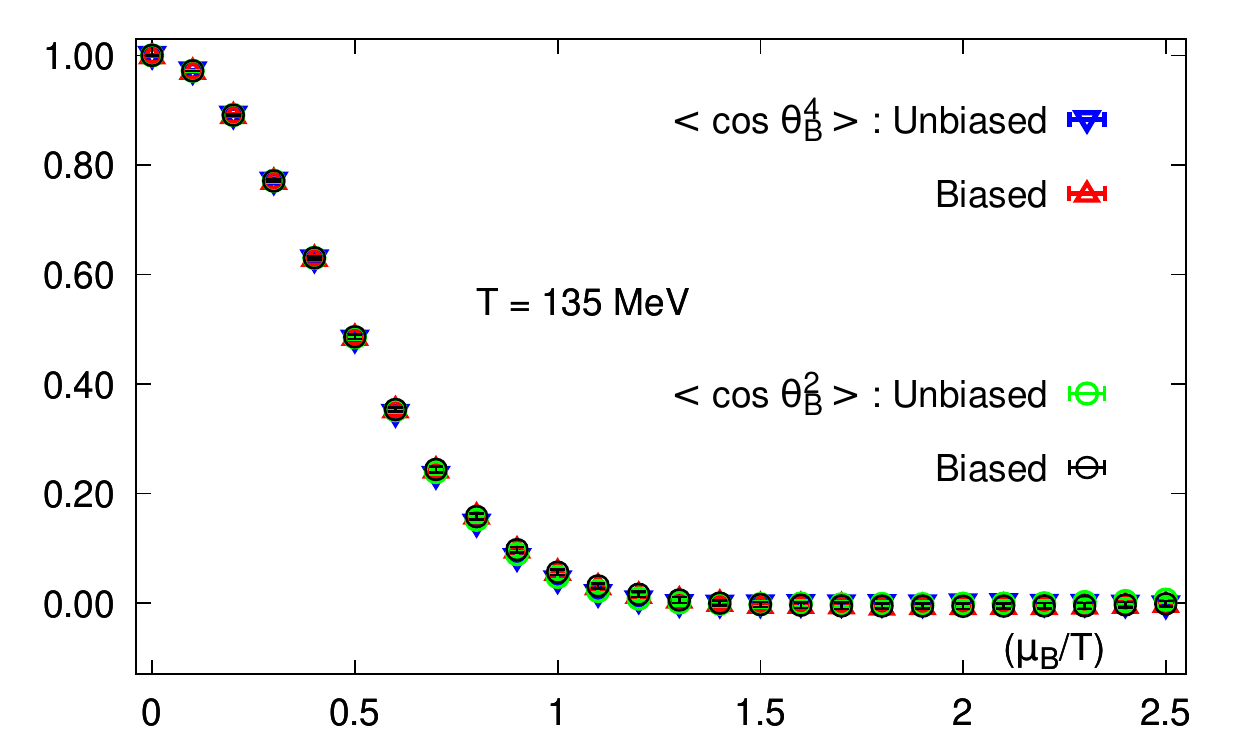} \\
    
    \includegraphics[width=0.47\textwidth]{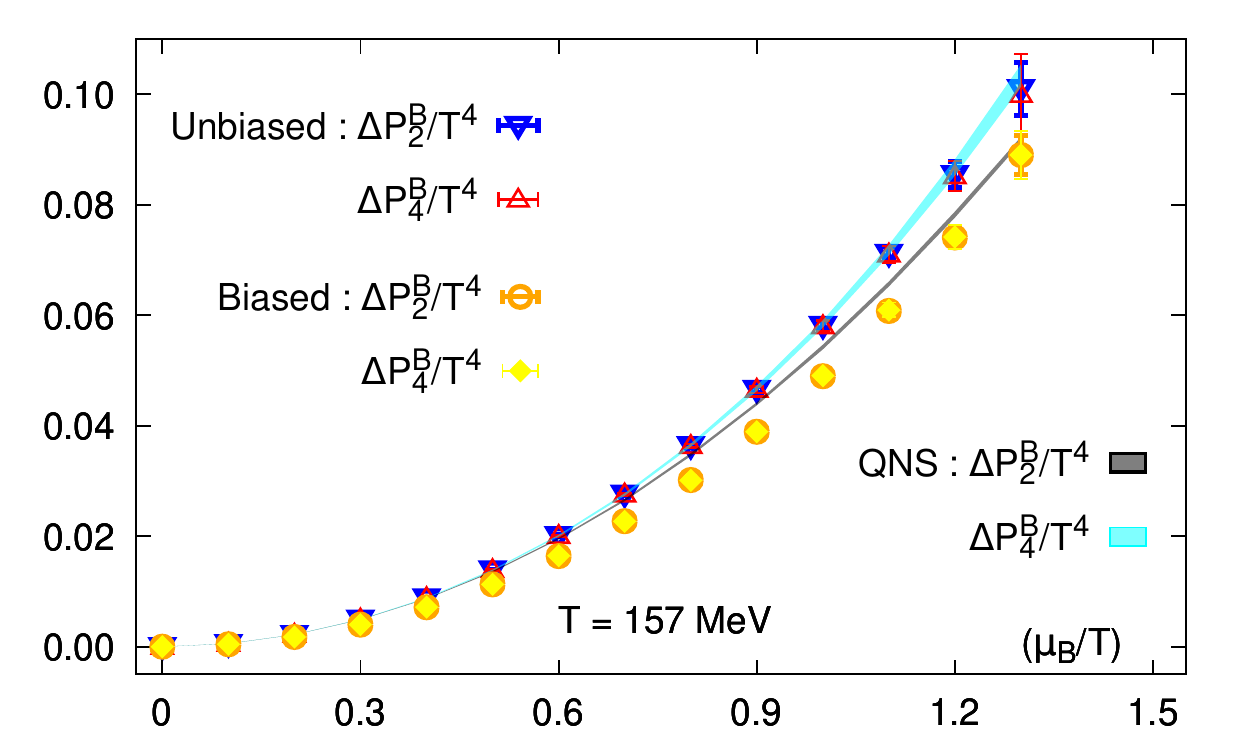} 
    \includegraphics[width=0.47\textwidth]{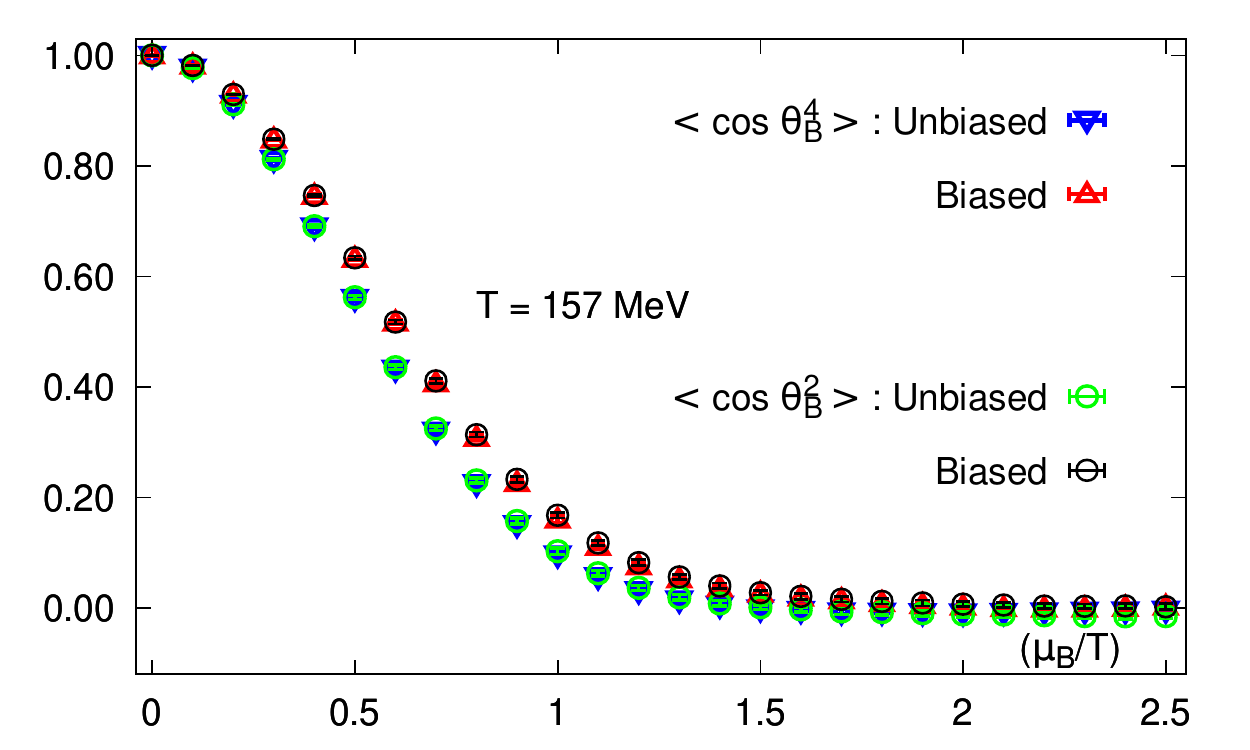}
 
    \caption{Baryon pressure (left) and phasefactor (right) plots for T = 135 and 157 MeV}
    \label{Fig:Baryon pressure}
\end{figure}

  The $2^{nd}$ and $4^{th}$ ordered unbiased pressure results in Fig.~\ref{Fig:Baryon pressure} are in better agreement with the $4^{th}$ ordered QNS results than the old biased counterparts and the difference is stark and highly pronounced for 135 MeV. Surely, one can argue for higher statistics reducing the gauge noise, allowing the comparison for higher values of $\mu_B/T$. Also, one can even vouch to increase the number of RVS from 500 to even more, per gauge configuration, specially for the noisiest $D_1$.   
  
  However, the solutions to these arguments come at the cost of huge computational time and storage space for data extraction of every $D_n$. The pressure plots demonstrate that even with a meagre $20K$ configurations with $\mathcal{O}(500)$ random vectors per configuration, the new formalism attains excellent agreement with QNS over the old one, thereby saving profound computational time and storage space. The imaginary part of the argument in the exponential function constitutes the phase-angle, the cosine of which forms the phasefactor in the biased and unbiased cases. The biased and unbiased phasefactor results in Fig.~\ref{Fig:Baryon pressure} vary slightly, besides showing appreciable order-by-order agreement, indicating that the difference between biased and unbiased pressure at 135 MeV is predominantly arising from phase quenched reweighting factor. 

 \section{Conclusions}

We have introduced a cumulant expansion which allows us to introspect the biased estimates and substitute them with unbiased counterparts order-by-order, in terms of $\mu_I$. The unbiased cumulant expansion, although truncated, managed to capture the higher-order fluctuations which the old exponential resummation could not efficiently serve to perform. Eventually, this results in the loss of reweighting factor and partition function $\mathcal{Z}$ itself. We then, therefore introduce a new exponential resummation formalism, which unlike the old resummation, exudes an excellent agreement with the QNS results, even using $20K$ configurations for $\mu_B$, with $\mathcal{O}(500)$ RVS per configuration. This enables to retrieve the partition function and hence, preserve the thermodynamics altogether. More significantly, this partially unbiased exponential resummation gives an all-ordered unbiased exponential resummation reproducing the exact all-ordered QNS in the limit of an infinite cumulant expansion series, apart from providing a much faster convergence with the QNS results. 

The unbiased exponential resummed approach, outlined here is a new way of extending the QCD EoS. Nevertheless, the possible connections between the approach presented here and various other proposals in the literature~\cite{Dimopoulos:2021vrk,Borsanyi:2021sxv,Pasztor:2020dur,Giordano:2020roi} still remain to be explored and therefore serve to be the promising ingredients for numerous future works.

\section*{Acknowledgments}

We sincerely thank all the members of the HotQCD collaboration for their inputs and for valuable discussions, as well as for allowing us to use their data from the Taylor expansion calculations. The computations in this work were performed on the GPU cluster at Bielefeld University, Germany. We thank the Bielefeld HPC.NRW team for their support.




 






\end{document}